\documentclass[12pt]{article}
\usepackage{amssymb, amsfonts, graphicx, hyperref, psfrag, epsfig, mathrsfs, color}
\newcommand{\be}{\begin{equation}}
\newcommand{\bea}{\begin{eqnarray}}
\newcommand{\eea}{\end{eqnarray}}
\newcommand{\ba}{\begin{array}}
\newcommand{\ea}{\end{array}}
\newcommand{\ee}{\end{equation}}
\newcommand{\pt}{\partial}

\csname @addtoreset\endcsname{equation}{section}
\begin{document}  
\begin{titlepage}

\title{\bf {Note on a noncritical holographic model with a magnetic field} \vspace{18pt}}
\vskip.3in

\author{\normalsize Sheng-liang Cui,$^{1}$ Yi-hong Gao,$^{1}$ Yunseok Seo,$^2$
  Sang-jin Sin$^3$ and Wei-shui Xu$^{2,3}$
  \vspace{12pt}\\
  ${}^1${\it\small  Key Laboratory of Frontiers in Theoretical Physics,}\\  
       {\it\small Institute of Theoretical Physics, Chinese Academy of Sciences,} \\
  {\it\small P.O. Box 2735,~Beijing 100190, ~China}\\
  ${}^2${\it\small Center for Quantum Spacetime, Sogang University, Seoul 121-742, Korea}\\
  ${}^3${\it\small Physics Department, Hanyang University, Seoul 133-791, Korea}\\
  {\small E-mail: { \it shlcui, gaoyh@itp.ac.cn, yseo, sjsin@hanyang.ac.kr,
      wsxuitp@gmail.com}} }

\date{}
\maketitle


\voffset -.2in \vskip 2cm \centerline{\bf Abstract} \vskip .4cm

We consider a noncritical holographic model constructed from an intersecting brane configuration
D4/$\overline{\rm{D4}}$-D4 with an external magnetic field. We investigate the
influences of this magnetic field on strongly coupled dynamics by the
gauge/gravity correspondence.

\vskip 4.0cm \noindent October 2009 \thispagestyle{empty}
\end{titlepage}

\newpage
\section{Introduction} 
The AdS/CFT correspondence \cite{Maldacena:1997re}-\cite{Aharony:1999ti} is a
useful method to study strongly coupled dynamics in gauge theory. In string
theory, some effective QCD-like theories can be constructed through intersecting
brane configurations. Then strongly coupled physics in gauge theories can be
investigated by the supergravity approximation. Recently, there are many
studies, such as \cite{Kruczenski:2003be}-\cite{Antonyan:2006vw}. For reviews, one can
see \cite{Mateos:2007ay}.

Through studying some holographic models in critical string
theory, we get some better understandings on strongly coupled physics in the
QCD-like effective theories. But there still exists many faults for the
critical holographic models. An important one is that color brane backgrounds
are ten-dimensional, so some part of such backgrounds need to be
compactified on some compact manifolds. It will produce some Kaluza-Klein(KK)
tower modes. However, in real QCD theory, there doesn't exist such KK
modes. Also some KK modes are at the same order as hadronic modes in the
QCD-like effective theory. So it is difficult to distinguish hadronic
modes from these KK modes. In order to overcome this point, one can consider
some intersecting brane configurations in noncritical string theory. The reason
is now gravity backgrounds lie at low dimension. In the Refs.
\cite{Kuperstein:2004yk}-\cite{Xu:2010}, such noncritical holographic
models were investigated, for example, the D4/D4-${\rm \overline{D4}}$ brane
configuration. However, the D-brane gravity backgrounds in noncritical
string theory have some shortcomings. The string coupling constants of these gravity
backgrounds are proportional to $1/N_c$. It means small string coupling constant
corresponds to large color number $N_c$. In the large $N_c$ limit, the 't Hooft
coupling constant $g_{YM}^2N_c$ is order one. The scalar curvature of gravity
background is also order one. Thus, the
gauge/gravity correspondence is not very reliable in noncritical string theory. 
But noncritical string models are still deserved to study. 
It will deepen our understandings on some
universal properties of general holographic models.

In \cite{Mazu:2007tp}, the authors consider an intersecting brane configuration,
which is composed of D4 and anti-D4 brane in six-dimensional noncritical string
theory. The color brane is D4, which extends along the directions $t, x_1,\cdots, x_4$.
The worldvolume coordinates of $N_f$ flavor D4-${\rm \overline{D4}}$ brane are
$t, x_1, \cdots, x_3$ and $u$. Under the quenched approximation $N_c\gg N_f$, the
backreaction of the flavor D4-${\rm \overline{D4}}$ on the color gravity
background can be omitted. Just like the Sakai-Sugimoto (SS) model
\cite{Sakai:2004cn}, we choose the coordinate $x_4$ to be periodic, then the
adjoint fermion on the color D4 brane satisfies an anti-periodic condition on the
$x_4$ circle. At low energy, they get mass and are decoupled. So the
final low energy effective theory on this intersecting brane configuration is a
four-dimensional QCD-like effective theory with a global chiral symmetry $U(N_f)_L\times
U(N_f)_R$ induced by $N_f$ D4-${\rm \overline {D4}}$ flavor brane pairs.
 
From the Refs. \cite{Kuperstein:2004yk}-\cite{Mazu:2007tp},
the near-horizon gravity background of D4 branes with a periodic coordinate
$x_4$ at low temperature is \bea
&&ds^2=\left(\frac{u}{R}\right)^2(~dt_E^2+dx_idx_i+f(u)dx_4^2~)+
\left(\frac{R}{u}\right)^2\frac{1}{f(u)}du^2, \nonumber \\
&&F_6=Q_c\left(\frac{u}{R}\right)^4dt\wedge dx_1\wedge dx_2\wedge
dx_3\wedge du\wedge dx_4,\label{background}\\
&&e^\phi=\frac{2\sqrt{2}}{\sqrt{3}Q_c},~~R^2=15/2,~~
f(u)=1-\left(\frac{u_{KK}}{u}\right)^5,\nonumber \eea where $i,~j=1,\cdots, 3$ and the parameter $Q_c$ is
proportional to the color brane number $N_c$. The Euclidean time is
periodic $t_E\sim t_E+\beta$. Since $\beta$ is arbitrary, the temperature
$1/\beta$ of this background is undetermined. In order to void a singularity, 
the coordinate $x_4$ needs to satisfy a periodic condition \be
x_4\sim x_4+\delta x_4=x_4+\frac{4\pi R^2}{5u_{KK}}.\label{periodica}\ee It
corresponds to a KK mass scale \be m_{KK}=\frac{2\pi}{\delta
  x_4}=\frac{5u_{KK}}{2R^2}.\label{kkscale}\ee

By a double wick rotation, we get a black hole solution. It reads \bea
&&ds^2=\left(\frac{u}{R}\right)^2(f(u)dt_E^2+dx_idx_i+
dx_4^2~)+\left(\frac{R}{u}\right)^2\frac{1}{f(u)}du^2,\nonumber\\
&&~~~~~~f(u)=1-\left(\frac{u_{T}}{u}\right)^5,
\label{blackbackground}\eea where the Euclidean time satisfies a periodic condition
\be t_E\sim t_E+\delta t_E=t_E+\frac{4\pi R^2}{5u_T},\label{periodicb}\ee and
now the radius of the coordinate $x_4$ is arbitrary. By comparing the free
energy between the gravity background (\ref{background}) and
(\ref{blackbackground}), we find there exists a first order Hawking-Page phase
transition (corresponding to the confinement/deconfinement phase transition in
the boundary theory) at critical temperature $\beta=\delta x_4$. Below this
temperature, the background (\ref{background}) is dominated. Otherwise, the
background (\ref{blackbackground}) will be dominated. These results are similar
to the cases \cite{Aharony:2006da} in the Sakai-Sugimoto model. From the gravity
backgrounds (\ref{background}) and (\ref{blackbackground}), it is clear that the 't
Hooft coupling constant is order one, and the curvature scalar is also order
one. Thus, it is not very reliable to use AdS/CFT
correspondence to study some strong coupled physics in this holographic model. 
In the following, we ignore this point and perform some investigations by using the usual method. 

In this paper, we consider to turn on an external magnetic field on the flavor
D4-${\rm \overline{D4}}$ branes just like the critical string cases
\cite{Albash:2006bs}-\cite{Seo:2009um}, and study its influences on strongly
coupled dynamics. Some effects of external magnetic field on the dynamics of QCD
theory were extensively studied in references, for example
\cite{Miransky:2002eb}. Here we investigate its influences in a noncritical
string model by the gauge/gravity correspondence. We choose this magnetic field
along the directions $x_2$ and $x_3$ on the worldvolume of D4 brane as follows \be 2\pi\alpha'
F_{23}=B.\label{magnetic}\ee Since there exists gauge invariance on the flavor
D4-brane worldvolume, this magnetic field is equivalent to a constant
Neveu-Schwarz--Neveu-Schwarz (NS-NS) field. So to investigate the D4-brane
classical dynamics with an external magnetic field becomes to study the flavor D4-brane
dynamics in the gravity background with a constant NS-NS field.

The organizations of this paper is as follows. In section two and three, we
investigate the flavor D4-brane dynamics in the low temperature background
(\ref{background}) and high temperature phase (\ref{blackbackground}),
respectively. In section four, we study a spinning fundamental string in the
gravity background (\ref{blackbackground}) and calculate the Regge trajectory
behaviors. The last section is a summary.

\section{Low temperature}
\begin{figure}[ht]
 \centering
 \includegraphics[width=0.5\textwidth]{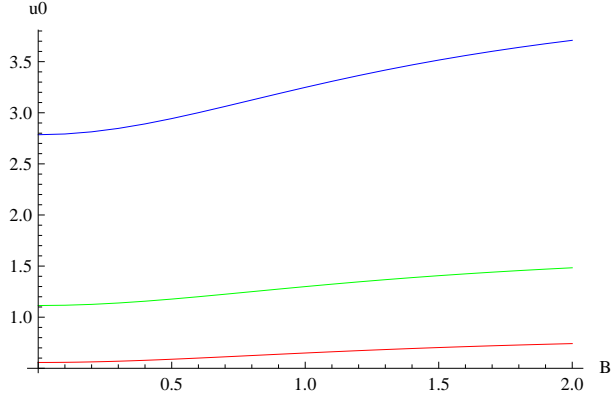}
 \caption{The connected point $u_0$ varies with the magnetic field $B$ at $L=1,~
   0.5$ and $ 
   0.2$ (from bottom to top). Here we choose $R=1$ and $u_{KK}=0.4$. }\label{figure1}
\end{figure}

In the low temperature phase, the gravity background is the equation
(\ref{background}). And we assume the worldvolume coordinate $u$ of flavor
D4-${\rm \overline{D4}}$ brane is depended on the background coordinate $x_4$. 
Then the induced metric on the worldvolume of flavor D4 brane is \be
ds^2=\frac{u^2}{R^2}(dt_E^2+\sum_{i=1}^{3}dx_idx_i)+\frac{u^2}{R^2}\left(f(u)(\pt_ux_4)^2+\frac{R^4}{f(u)u^4}\right)du^2.\ee
With the magnetic field (\ref{magnetic}), the DBI action for the flavor D4 brane is \footnote{Follow the arguments in
\cite{Mazu:2007tp}, here we don't consider the contribution of the Chern-Simons (CS) term.} \be S\sim \int
du\frac{u^3}{R^3}\sqrt{\left(\frac{u^4}{R^4}+B^2\right)\left(f(u)(\pt_ux_4)^2+\frac{R^4}{f(u)u^4}\right)}.
\label{action1}
\ee So the equation of motion is derived as \be \frac{\pt}{\pt{x_4}}\left(
  \frac{u^5}{R^5}\frac{f(u)\sqrt{1+B^2\frac{R^4}{u^4}}}{\sqrt{f(u)+\frac{R^4}{u^4f(u)}(\pt_{x_4}u)^2}}\right)=0.\label{eom1}\ee
We choose a boundary condition as $u'=0$ at $u=u_0$ (where $'=\pt_{x_4}$). It
means $u_0$ is a connected point between the flavor D4 and anti-D4
branes. After an integration, the equation (\ref{eom1}) becomes \be
\frac{u^5}{R^5}\left(\frac{f(u)\sqrt{1+B^2\frac{R^4}{u^4}}}{\sqrt{f(u)+\frac{R^4}{u^4f(u)}(\pt_{x_4}u)^2}}\right)=\frac{u_0^5}{R^5}\sqrt{f(u_0)(1+B^2\frac{R^4}{u_0^4})}. \ee
Define $y\equiv \frac{u}{u_0}$, $y_{KK}\equiv \frac{u_{KK}}{u_0}$ and
$f(y)= 1-\frac{y_{KK}^5}{y^5}$, we get \be
u'=\frac{u^2f(y)}{R^2}\sqrt{\frac{f(y)(1+B^2\frac{R^4}{u_0^4}y^{-4})y^{10}}{(1+B^2\frac{R^4}{u_0^4})f(1)}-1}. \label{solution1}\ee
Then the asymptotic distance between the D4 and anti-D4 brane is \bea
L&=&2\int_{u_0}^\infty \frac{du}{u'}\cr
&=&\frac{2R^2}{5u_0}\int_0^1dz\frac{z^{1/5}\sqrt{(1-y_{KK}^5)(1+B^2\frac{R^4}{u_0^4})}}{(1-y_{KK}^5z)\sqrt{(1+B^2\frac{R^4}{u_0^4}z^{4/5})(1-y_{KK}^5z)-z^2(1-y_{KK}^5)(1+B^2\frac{R^4}{u_0^4})}},
\eea where $z=y^{-5}$. So the
connected point $u_0$ satisfies the equation \be
u_0=\frac{2R^2}{5L}\int_0^1dz
\frac{z^{1/5}\sqrt{(1-y_{KK}^5)(1+B^2\frac{R^4}{u_0^4})}}{(1-y_{KK}^5z)
\sqrt{(1+B^2\frac{R^4}{u_0^4}z^{4/5})(1-y_{KK}^5z)-z^2(1-y_{KK}^5)(1+B^2\frac{R^4}{u_0^4})}}.\label{uo1}\ee
So $u_0$ depends on the parameters $B$ and $L$. Its dependence is plotted in Fig.~\ref{figure1}. 
It shows that the joint point increases with increasing the magnetic field $B$,
and decreases as the distance $L$ increases.

.

By inserting the equation (\ref{solution1}) to the action (\ref{action1}) ,
the on-shell action of the connected solution is \bea
S_{connected}&\sim&\int_1^\infty dy
\frac{y^3(1+B^2\frac{R^4}{u_0^4}y^{-4})}{\sqrt{(1+B^2\frac{R^4}{u_0^4}y^{-4})f(y)
-(1+B^2\frac{R^4}{u_0^4})f(1)y^{-10}}}\cr
&\sim&
\int_0^1dz\frac{1}{z^{9/5}}\frac{1+B^2\frac{R^4}{u_0^4}z^{4/5}}{\sqrt{(1+
B^2\frac{R^4}{u_0^4}z^{4/5})(1-y_{KK}^5z)-z^2(1-y_{KK}^5)(1+B^2\frac{R^4}{u_0^4})}}. \eea
In the gravity background (\ref{background}), there doesn't exist separated
flavor D4 and ${\rm \overline{D4}}$ solution. The reason is the flavor branes
don't have any place to end in this background. If the coordinate $x^4$ is not
periodic, and there is not $f(u)$ factor in this background, then the separated
flavor solution will be existed \cite{Antonyan:2006vw}. Thus, the global chiral
symmetry $U(N_f)_L\times U(N_f)_R$ is always broken to its diagonal part at low
temperature.

\begin{figure}[ht]
 \centering
 \includegraphics[width=0.5\textwidth]{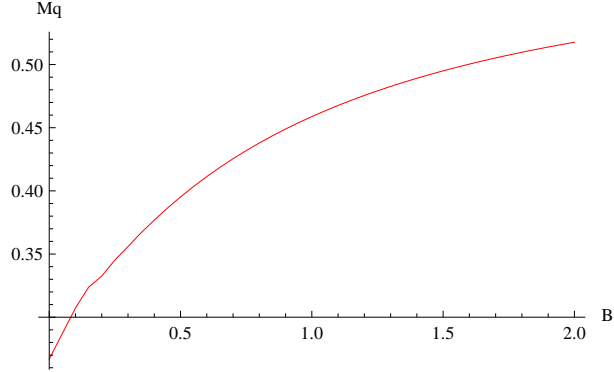}
 \caption{It shows the dependence of the quark condensation on the magnetic
   field $B$ at $u_{KK}=0.4$. }
 \label{figure2}
\end{figure}

This connected solution corresponds to the chiral symmetry breaking phase in
the gauge theory side. It means there exists a quark condensation. Its energy
scale corresponds to the length of a fundamental string connected between $u_{KK}$ and $u_0$
in the background (\ref{background}). It reads \be
M_q=\frac{1}{2\pi\alpha'}\int_{u_{KK}}^{u_0}du
\sqrt{g_{{t_E}{t_E}}g_{uu}}=\frac{1}{2\pi\alpha'}\int_{u_{KK}}^{u_0}\frac{du}{\sqrt{f(u)}}=
\frac{u_0}{2\pi\alpha'}\int_{y_{KK}}^1\frac{dy}{\sqrt{f(y)}}.\label{mq1}\ee 
By inserting the equation (\ref{uo1}) for $u_0$ into the equation (\ref{mq1}),
we get 
\bea M_q&=&\frac{2R^2}{10L\alpha'}\int_{y_{KK}}^1\frac{dy}{\sqrt{f(y)}}\cr
&&\cdot\int_0^1dz
\frac{z^{1/5}\sqrt{(1-y_{KK}^5)(1+B^2\frac{R^4}{u_0^4})}}{(1-y_{KK}^5z)
\sqrt{(1+B^2\frac{R^4}{u_0^4}z^{4/5})(1-y_{KK}^5z)-z^2(1-y_{KK}^5)(1+B^2\frac{R^4}{u_0^4})}}. \eea
For a fixed asymptotic distance $L$, we plot the Fig.~\ref{figure2} by using the
same numerical way in \cite{Johnson:2008vna}.
From this figure, the condensation energy scale increases with increasing the
magnetic field $B$. Its behavior almost grows like $B^{1/6}$. And the scale
of chiral symmetry breaking becomes large with increasing the magnetic field
$B$.

\section{high temperature}
In the high temperature background (\ref{blackbackground}), by using the same
embedding ansatz as the low temperature case, the induced metric on the
flavor D4-brane is \be
ds^2=\frac{u^2}{R^2}(f(u)dt^2+\sum_{i=1}^3dx_i^2)+\frac{u^2}{R^2}\left(\left(\frac{\pt
      x_4}{\pt u}\right)^2+\frac{R^4}{u^4f(u)}\right)du^2.\ee Then the D4-brane
effective action is \be S\sim \int dx_4
\frac{u^5}{R^5}\sqrt{(1+B^2\frac{R^4}{u^4})(f(u)+\frac{R^4}{u^4}u'^2)}.\label{action2}\ee
And the equation of motion is \be
\frac{\pt}{\pt{x_4}}\left(\frac{u^5}{R^5}\frac{f(u)\sqrt{1+B^2\frac{R^4}{u^4}}}
  {\sqrt{f(u)+\frac{R^4}{u^4}u'^2}}\right)=0.\ee Like as the low temperature
case, we choose a boundary condition $u'=0$ at $u=u_0$. Then we get a first
derivative equation of motion \be
\frac{u^5}{R^5}\frac{f(u)\sqrt{1+B^2\frac{R^4}{u^4}}}{\sqrt{f(u)+\frac{R^4}{u^4}u'^2}}
=\frac{u_0^5}{R^5}\sqrt{f(u_0)(1+B^2\frac{R^4}{u_0^4})}. \label{connectsolutions}\ee
With the definition $y\equiv \frac{u}{u_0}$, the above equation becomes \be
y'=u_0\frac{y^2}{R^2}\sqrt{f(y)}\sqrt{\frac{(1+B^2\frac{R^4}{u_0^4}y^{-4})
    y^{10}f(y)}{(1+B^2\frac{R^4}{u_0^4})f(1)}-1}.\ee

\begin{figure}[ht]
 \centering
 \includegraphics[width=0.5\textwidth]{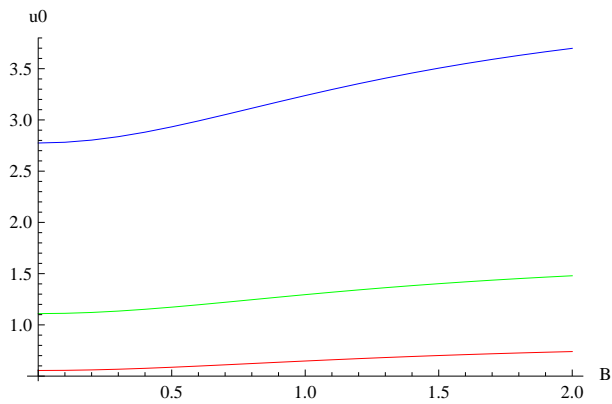}
 \caption{The joint point $u_0$ depends on the magnetic field $B$ at
   $L=1,~ 0.5$ and $0.2$ (from bottom to top). We set $u_T=0.3$ and $R=1$. } \label{figure3}
\end{figure}

Similarly, the asymptotic distance $L$ between the D4 and anti-D4 brane
reads \be L=2\int_{u_0}^\infty \frac{du}{u'} =\frac{u_0}{R^2}\int_1^\infty dy
\frac{\sqrt{(1+B^2\frac{R^4}{u_0^4})f(1)}}{y^2\sqrt{f(y)}
\sqrt{(1+B^2\frac{R^4}{u_0^4}y^{-4})y^{10}f(y)-(1+B^2\frac{R^4}{u_0^4})f(1)}}.\ee 
It can be written as \bea
&&u_0=\frac{2R^2}{5L}\sqrt{(1+B^2\frac{R^4}{u_0^4})(1-y_T^5)}\cr 
&&\cdot\int_0^1dz\frac{z^{1/5}}{\sqrt{(1-y_T^5)\left((1+B^2\frac{R^4}{u_0^4}z^{4/5})(1-y_T^5z)
-z^2(1+B^2\frac{R^4}{u_0^4})(1-y_T^5)\right)}},\label{uo2}
\eea 
Some numerical results of $u_0(B, L)$ are shown in Fig.~\ref{figure3}.
It is clear that the point $u_0$ decreases by increasing the asymptotic distance $L$. And this
point increases as the magnetic field $B$ increases. These results are similar to
the cases at zero temperature.

 
After substituting the equation (\ref{connectsolutions}) into the action
(\ref{action2}), we get the
on-shell energy of the connected D4-$\rm\overline{D4}$ brane solution as follows
 \be S_{\rm connected}\sim \int_1^\infty dy
\frac{y^3(1+B^2\frac{R^4}{u_0^4}y^{-4})\sqrt{f(y)}}{\sqrt{(1+B^2\frac{R^4}{u_0^4}y^{-4})f(y)
-(1+B^2\frac{R^4}{u_0^4})f(1)y^{-10}}}. \ee
Now the gravity background is a black hole background (\ref{background}), so
there exists a separated D4 and anti-D4 brane solution $u'\rightarrow \infty$. Its
on-shell energy is \be S_{\rm separated}\sim \int_0^\infty dy
y^3\sqrt{1+B^2\frac{R^4}{u_0^4}y^{-4}}. \ee Then the energy difference between
the connected and separated solution is \bea \delta S &\sim& \int_1^\infty dy
\left(\frac{y^3(1+B^2\frac{R^4}{u_0^4}y^{-4})\sqrt{f(y)}}{\sqrt{(1+B^2\frac{R^4}{u_0^4}y^{-4})f(y)
-(1+B^2\frac{R^4}{u_0^4})f(1)y^{-10}}}-y^3\sqrt{1+B^2\frac{R^4}{u_0^4}y^{-4}}\right)\cr
&&~~~~~~~~-\int_{y_T}^1 dy y^3\sqrt{1+B^2\frac{R^4}{u_0^4}y^{-4}}.\eea
\begin{figure}[ht]
 \centering
 \includegraphics[width=0.6\textwidth]{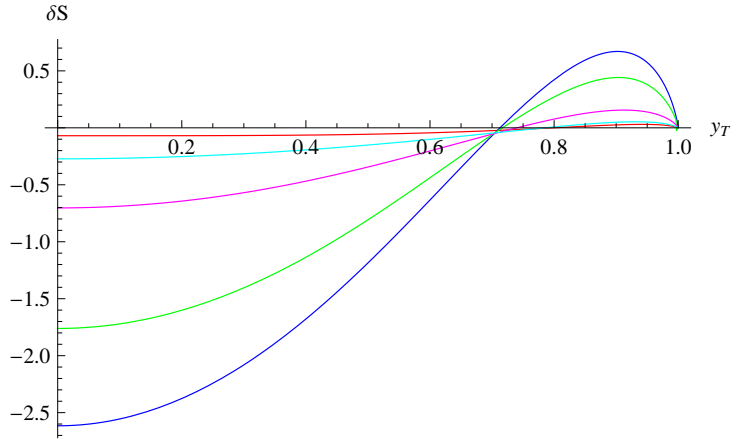}
 \caption{The energy difference depends on $y_T$ at different values $B=0,~ 1,~
   3,~ 8$ and $12$ (from
   red to blue, or the length of dashed line segment is increased).  } \label{figure4}
\end{figure} Its numerical result is shown in Fig.~\ref{figure4}. The 
energy difference has two branches. Below some critical temperatures,
the difference is negative. Now the connected solution is dominated, and means that
the chiral symmetry in the gauge theory is broken. Above this critical temperature, the energy
difference is positive, the separated solution is dominated and the chiral
symmetry is restored. Also the critical point $y_T$ decreases with increasing
the magnetic field $B$.
\begin{figure}[ht]
  \centering
 \includegraphics[width=0.5\textwidth]{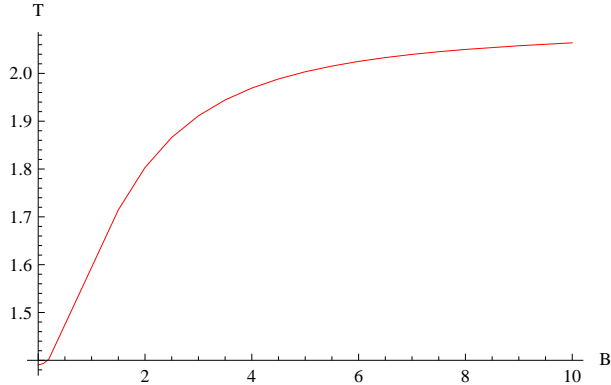}
 \caption{The critical temperature of chiral phase transition depends on the magnetic field $B$. }\label{figure5}
\end{figure}
In the unit of $1/L$, we draw the Fig.~\ref{figure5}, which shows how the
critical temperature to vary with the magnetic field $B$. Above this curve, it
denotes the chiral restoration phase. Below it, this is the chiral symmetry
breaking phase. The critical
temperature of chiral symmetry restoration increases as the magnetic
field $B$ increases. Here these results are also 
similar to some results in \cite{Johnson:2008vna} and \cite{Seo:2009um}. It is also
consist with some already known results in field theory with a magnetic
background field \cite{Miransky:2002eb}.

\begin{figure}[ht]
 \centering
 \includegraphics[width=0.5\textwidth]{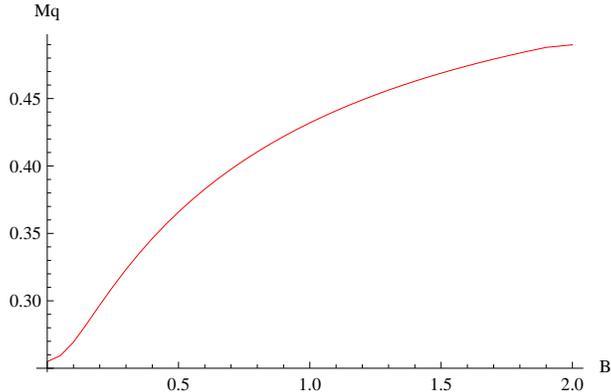}
 \caption{The quark condensation $M_q$ varies with the magnetic field
   $B$ at $u_T=0.3$. }\label{figure6}
\end{figure}

In the chiral symmetry broken phase, there exists a quark condensation in gauge
theory side. The condensation energy scale corresponds to the string length
between the connected point $u_0$ and the horizon of black hole. By using
the equation (\ref{uo2}), the energy scale of quark condensation is derived as \be
M_q=\frac{1}{2\pi\alpha'}\int_{u_T}^{u_0}du=\frac{u_0}{2\pi\alpha'}(1-y_T).\label{mq2} \ee
Its dependence on the magnetic field $B$ is plotted in Fig.~\ref{figure6}.
Just like the low temperature cases, now the condensation also
increases with increasing the magnetic field $B$. When the magnetic field is
located around the region $(0, 0.5)$, its dependence is almost linear. But above
this region, the dependence is similar to the low temperature case. From
the equation (\ref{mq2}), this quark condensation vanishes at $y_T=1$. It
corresponds to a chiral phase transition point.

\section{Quark and Meson in hot QGP}
In the introduction, we already discussed that turn on a magnetic field on the
flavor D4-brane is equivalent to add a NS-NS field into the gravity backgrounds
(\ref{background}) and (\ref{blackbackground}). From the supergravity action of
noncritical string action \cite{Kuperstein:2004yk}, this new background is
still a solution. In the following, we mainly consider a fundamental sting in
this new gravity background with a NS-NS field (\ref{magnetic}). 

Firstly, we consider a quark moving into the hot plasma. The quark corresponds
to one endpoint of fundamental string on the flavor brane. By using the method in
\cite{Gubser:2006bz}, we parametrize the world-sheet coordinates of this
fundamental string as $\tau=t$ and $\sigma=u$, and assume the endpoint (quark)
on the flavor brane moving along the direction $x_2$ with \be x_2=vt+\xi(u),\ee
where $v$ is the velocity. Because of a rotational symmetry in $x_2$ and
$x_3$ plane, it is equivalent to let quark move along the direction $x_3$. It is
easy to see the Wess-Zumino term in the string action vanishes. So this NS-NS
field (\ref{magnetic}) doesn't produce any contributions on the drag force and
energy loss for quark moving through this hot plasma. But in the non-trivial
NS-NS background, the influence of the NS-NS background field is investigated in
\cite{Seo:2009um} and \cite{Matsuo:2006ws}. Usually, a NS-NS background field
will decrease the drag force and energy loss of a quark moving through the
hot-QGP (quark-gluon-plasma).

Now we turn to consider high spin mesons. We mainly focus on the chiral
symmetry breaking phase at high temperature. Now the flavor D4 and ${\rm
  \overline{D4}}$ are connected each other through a wormhole. The bound state
of two endpoints of a spinning fundamental string on the flavor D4-${\rm \overline{D4}}$
brane pairs corresponds to a high spin meson in the boundary effective
theory. Define $\rho^2=x_2^2+x_3^2$,  we rewrite the background
(\ref{blackbackground}) as \bea
&&ds^2=\left(\frac{u}{R}\right)^2\left(f(u)dt_E^2+dx_1^2+d\rho^2+\rho^2
  d\phi^2+dx_4^2\right)+\left(\frac{R}{u}\right)^2\frac{1}{f(u)}du^2, \cr &&~~
f(u)=1-\left(\frac{u_T}{u}\right)^5, \eea and the NS-NS background field is \be
B dx_2\wedge dx_3=B\rho d\rho\wedge d\phi.\ee

We choose the string worldsheet coordinates as \be \tau=t_E, ~~\sigma=\rho,
~~u(\sigma),~~ \phi=\omega\tau. \ee So the Nambu-Goto action of this spinning fundamental
string is \be S_{NG}=-\frac{1}{2\pi\alpha'}\int d\tau d\sigma
\left(\frac{u}{R}\right)^2\sqrt{(f(u)-\rho^2\omega^2)\left(1+\frac{1}{f(u)}
\left(\frac{R}{u}\right)^4u'^2\right)}+\frac{1}{2\pi\alpha'}\int
d\tau d\sigma B\rho\omega, \label{stringaction}\ee where $'=\pt_\rho$. Then we
can obtain the equations of motion for $u$. For simplicity, we don't show those
equations here. Set the boundary conditions as $u'\rightarrow \infty$ at the boundary
and $u'=0$ at $u=u_0$, we plot the shape of this spinning string in the
Fig.~\ref{shape1} and Fig.~\ref{shape2} (Here we only plot the zero-node
cases. )\footnote{In plotting all the following figures, we choose $u_0=20$ and
  $R=u_T=1$.}.
\begin{figure}[ht]
 \centering
 \includegraphics[width=0.55\textwidth]{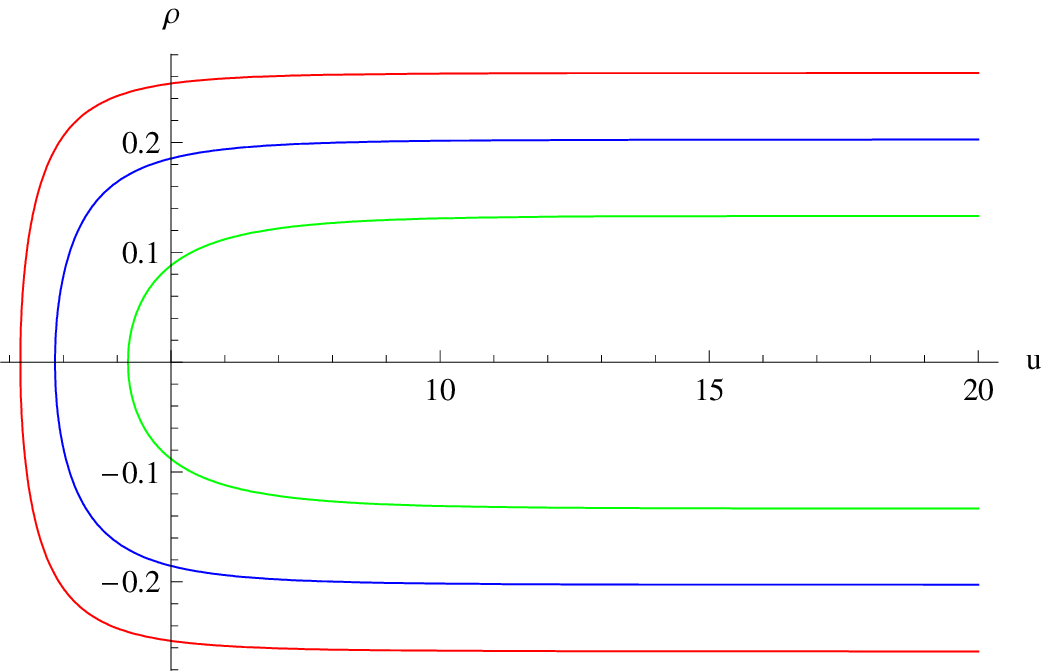} 
 \caption{It is the string shape at $B=0$ and $w=1,~ 1.5$ and $3$ (red, blue and
   green) from left to right. } \label{shape1}
 \begin{tabular}{cc}
 \includegraphics[width=0.47\textwidth]{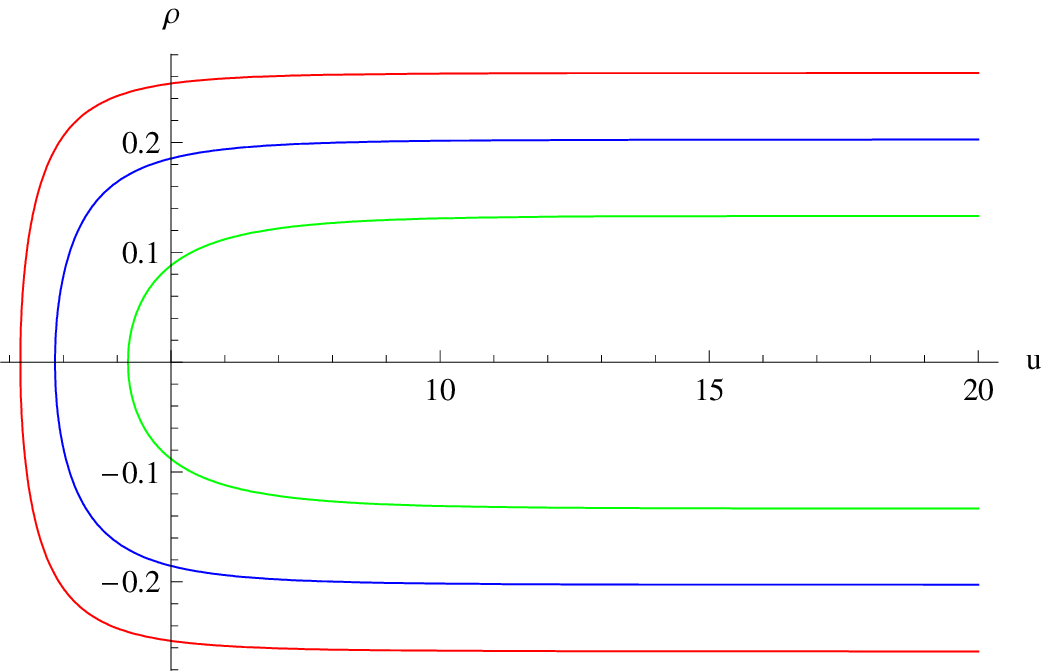}&
 \includegraphics[width=0.47\textwidth]{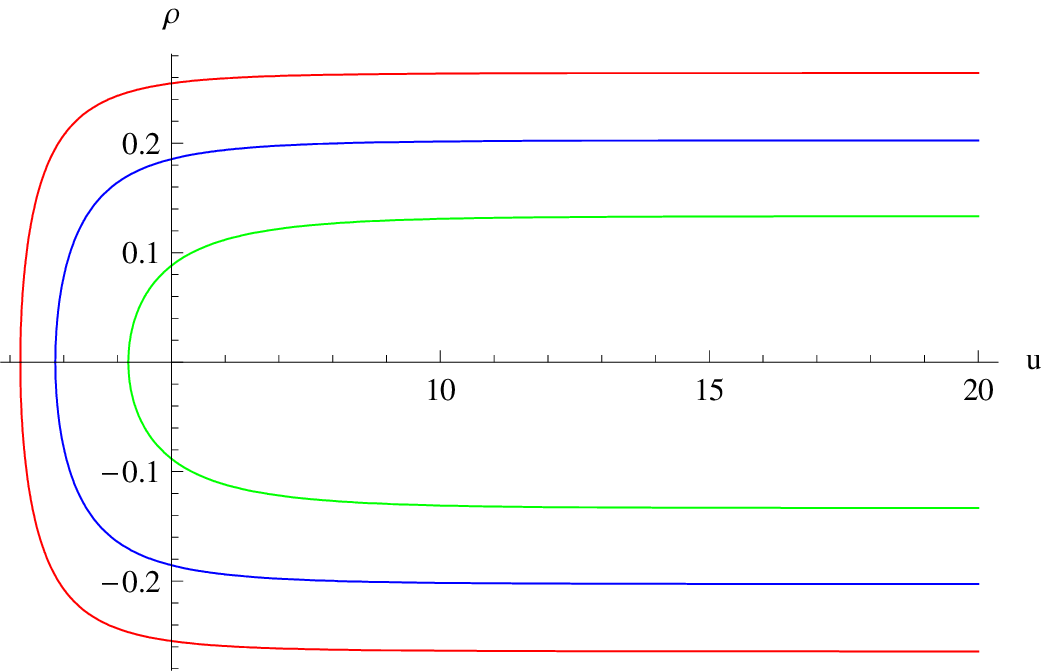}\\
  (a) & (b)
 \end{tabular}
  \caption{It is the string shape at $w=1,~ 1.5$ and $3$ (red, blue and
   green) from left to right. (a) $B=3$;  (b) $B=5$. } \label{shape2}
\end{figure} It shows the asymptotic distance for two
endpoints of fundamental string decreases as the angular
velocity $\omega$ increases. And the turning point of fundamental string becomes
large with increasing this angular velocity. But the influences of the NS-NS
field on the string shape are not very sensitive.

From the string action (\ref{stringaction}), it is clear there exists two
conserved quantities for this spinning string. One is the energy $E$, the other
one is the angular momentum $J$. Their expressions are derived as \bea &&
E=\frac{1}{2\pi\alpha'}\int d\tau d\sigma
\left(\frac{u}{R}\right)^2\frac{f(u)\sqrt{1+\frac{1}{f(u)}
    \left(\frac{R}{u}\right)^4u'^2}}{\sqrt{f(u)-\rho^2\omega^2}},\label{energy}\\
&& J=\frac{1}{2\pi\alpha'}\int d\tau d\sigma~
\omega\rho^2\left(\frac{u}{R}\right)^2\frac{\sqrt{1+\frac{1}{f(u)}\left(\frac{R}{u}\right)^4u'^2}}
{\sqrt{f(u)-\rho^2\omega^2}}+\frac{1}{2\pi\alpha'}\int d\tau d\sigma
B\rho.\label{angular} \eea

By doing some numerical calculations, we
plot some figures to show the relations $E^2(\omega)$, $J(\omega)$ and $E^2(J)$.
\begin{figure}[ht]
 \centering
 \begin{tabular}{cc}
 \includegraphics[width=0.45\textwidth]{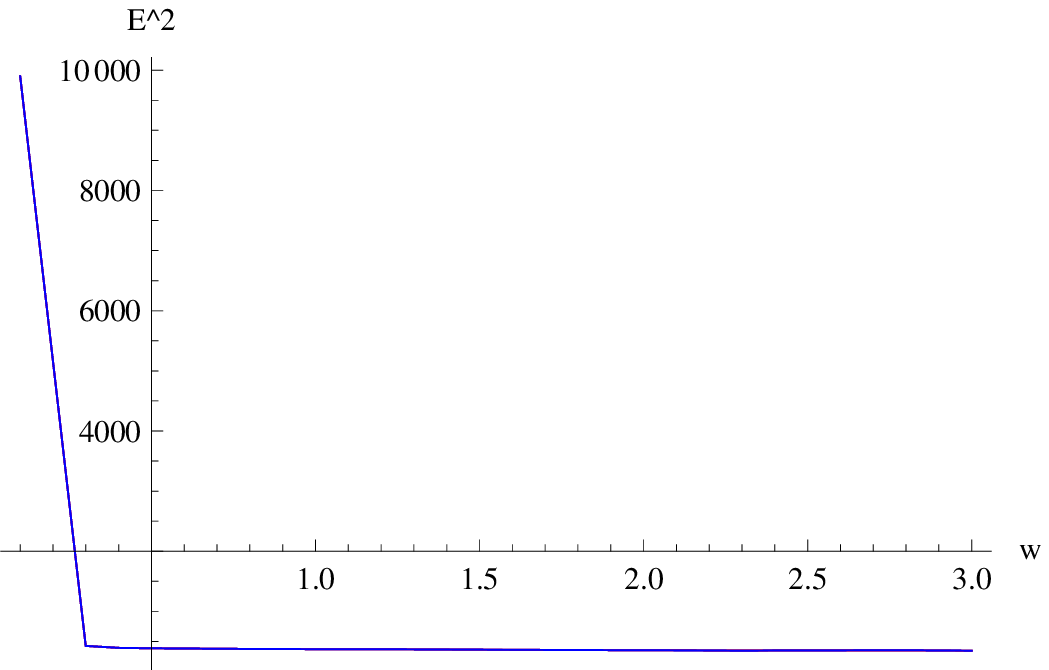}&
 \includegraphics[width=0.45\textwidth]{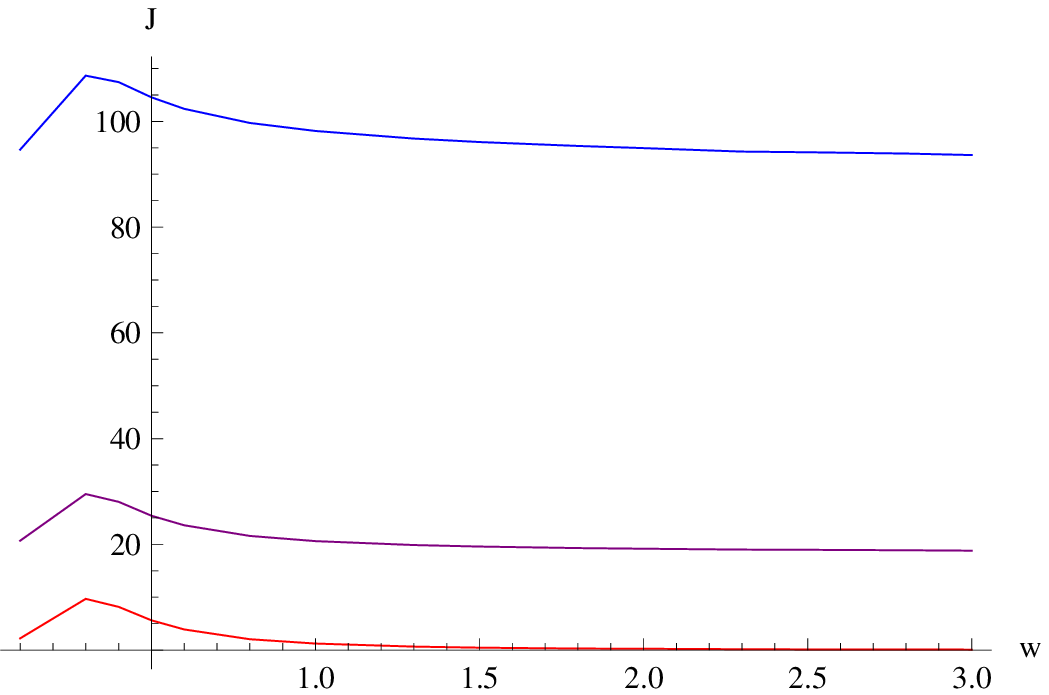}\\
  (a) & (b)
 \end{tabular}
 \caption{(a) $E^2$ varies with the angular velocity $\omega$ at $B=0,~ 3$ and
   $5$. These three curves are totally overlapped;  (b) angular
   momentum $J$ varies with the angular velocity $\omega$ at $B=0,~ 3$ and $5$ (red, purple,
   blue) from bottom to top. } \label{figure9}
\includegraphics[width=0.53\textwidth]{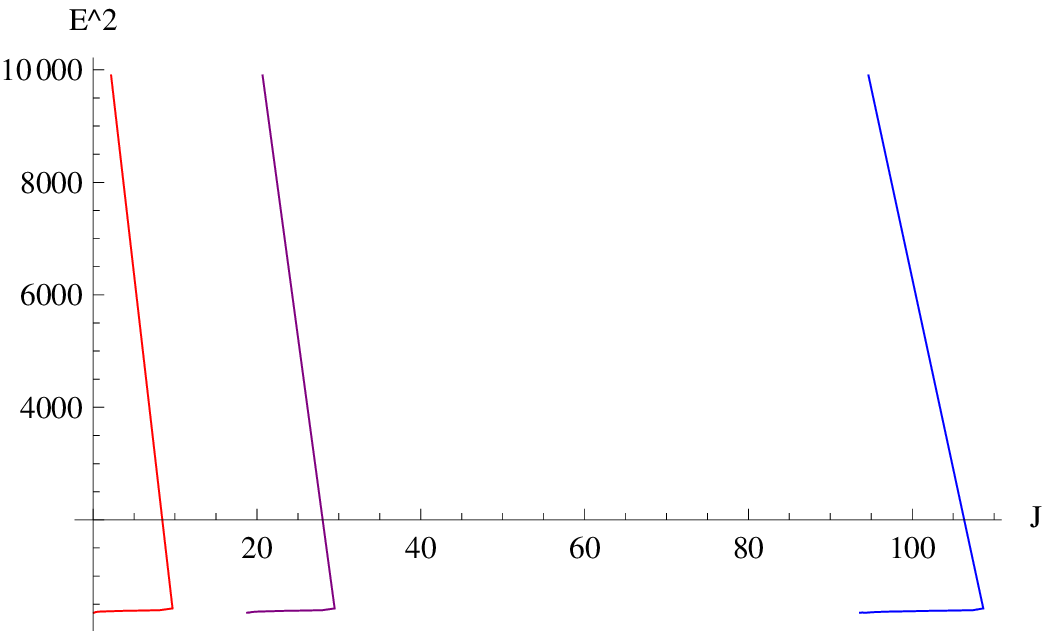}
 \caption{The energy $E^2$ varies with the angular momentum $J$ at $B=0,~3$ and
   $5$  from left to right (red, purple, blue). } \label{figure10}
\end{figure}
The energy $E^2$ decreases as the angular velocity $\omega$ increases, and
it is not sensitive to different NS-NS background field value $B$. For all the
different magnetic field $B$, the Fig.~\ref{figure9}(a) shows that the
energy of spinning string has a similar dependence on the angular velocity
$\omega$. From the Fig.~\ref{figure9}(b), the angular momentum $J$ have
maximums at some particular points $\omega$. As the angular velocity
increases, the angular momentum $J$ also increases. However, after maximum
points, it will decreases. For a larger NS-NS field $B$, the angular momentum
$J$ becomes large. In Fig.~\ref{figure10}, it is the Regge trajectory
behavior $E^2(J)$, which has two branches relative to the angular
momentum $J$. With increasing the angular momentum (increasing the angular
velocity), the energy square $E^2$ decreases. Beyond maximums, the value
$E^2$ also decreases as the angular momentum $J$ (still increasing the angular
velocity) decreases. These behaviors have some differences with some results in some critical
string holographic models \cite{Peeters:2006iu}, \cite{Johnson:2009ev} and
\cite{Seo:2009um}.

\section{Summaries}
In this paper, we mainly consider a noncritical string holographic model with an
external magnetic background field. We investigate the influences of this
magnetic field on the underlying dynamics by using the gauge/gravity
correspondence. In section two and three, we mainly consider the chiral symmetry
breaking in low temperature and high temperature phase. The scale of the chiral
symmetry breaking increases as the magnetic field increases. At high
temperature, the critical temperature of the chiral phase transition is
different at a different magnetic field. In the unit of $1/L$, this phase
transition temperature increases as the magnetic field increases. Finally, we
investigate high spin mesons in the chiral symmetry broken phase at high
temperature. Our results here, except for the Regge behavior, are similar to the
cases in some other critical string holographic model. It confirms some
universal properties of holographic models constructed through intersecting
brane configurations in string theory.

\subsection*{Acknowledgments}
This work of Yunseok was supported by the Korea Science and Engineering
Foundation (KOSEF) grant funded by the Korea government(MEST) through the Center
for Quantum Spacetime(CQUeST) of Sogang University with grant number
R11-2005-021. The work of SJS and Weishui is supported by KOSEF Grant
R01-2007-000-10214-0 and also by the SRC Program of the KOSEF through the Center
for Quantum SpaceTime (CQUeST) of Sogang University with grant number
R11-2005-021.

\end{document}